\documentclass[runningheads]{llncs}
\usepackage{graphicx}
\usepackage[lofdepth,lotdepth]{subfig}
\usepackage{lipsum}
\usepackage{multicol}
\newcommand{\hlhref}[2]{\href{#1}{\textcolor{blue}{\underline{#2}}}}

\usepackage{ulem}
\usepackage[english]{babel}

\usepackage{fancyhdr} %add page number
\usepackage{float}

\usepackage{pifont} % use command \ding
\usepackage{amssymb}
\usepackage{bbding} %对勾叉号
\usepackage{pifont}

\usepackage{makecell}
\usepackage{colortbl} %table color

\usepackage{wrapfig} % 浮动表格
\usepackage[ruled]{algorithm}
\usepackage{algorithmic}
\usepackage{siunitx}
\usepackage{longtable}
\usepackage{tikz}
\usetikzlibrary{trees}
\usepackage{array}
\newcolumntype{L}[1]{>{\raggedright\arraybackslash}p{#1}}

\usepackage{url}

\usepackage{hyperref}
\hypersetup{
    colorlinks=true,
    linkcolor=blue,
    filecolor=magenta,      
    urlcolor=magenta,
    citecolor = magenta,
    pdftitle={An Example},
    pdfpagemode=FullScreen,
    }
\usepackage{tikz} % to be able to draw some self-contained figs
\usepackage{comment}
\usepackage{filecontents}
\usepackage{multirow}

\usepackage{amsmath,amssymb,amsfonts,amsthm}
\usepackage{graphicx}
\usepackage{textcomp}
\usepackage{xcolor}
\usepackage{amsthm}
\usepackage{booktabs}
\usepackage[utf8]{inputenc}

\usepackage{dashbox}
\usepackage{todonotes}
\usepackage{enumitem}   
\usetikzlibrary{shapes.callouts}
\usepackage{listings}
\usepackage{trace}
\usepackage{makeidx}
\usepackage{mdframed}

\theoremstyle{plain}

\theoremstyle{definition}

\usepackage{threeparttable} %制作三线表格
\tikzset{%
    pics/sema/.style args={#1/#2/#3}{code={%
        \ifstrequal{#2}{0}{%
            \node[circle,minimum width=1mm,draw,fill=#1] {};
        }{%
            \tkzDefPoint(0,0){O}
            \tkzDrawSector[R,fill=#1](O,1mm)(90,90-#2)
            \tkzDrawSector[R,fill=#3](O,1mm)(90-#2,90-360)
    }
    }},
}

\usepackage{dashbox}
\usepackage{tikz}

\usepackage{ulem}

\begin{document}
%====================================================
%\title{Advancing Beyond ORDI: The Next Frontier in Bitcoin Inscriptions Technology}
\title{Bitcoin Inscriptions: Foundations and Beyond}
\subtitle{(Tech Report)}

%====================================================
%====================================================
\author{Ningran Li\inst{1} \and
Minfeng Qi\inst{1} \and
Qin Wang\inst{2} \and
Shiping Chen\inst{2} }
\authorrunning{N. Li et al.}
\institute{
Swinburne University of Technology, Australia
%\email{lncs@springer.com}
\\ 
\and
CISRO Data61, Australia\\
}

%====================================================
\maketitle             

\begin{abstract}
Bitcoin inscription marks a pivotal moment in blockchain technology. This report presents a primary exploration of Bitcoin inscriptions. We dive into the technological underpinnings and offer a detailed comparative analysis between Bitcoin inscriptions and NFTs on other blockchains. Further, we explore a wide range of use cases and significant opportunities for future innovation, including inscription derivative protocols, Bitcoin Layer2 solutions, and interoperability techniques. 

\end{abstract}

%====================================================
%====================================================

\section{Bitcoin New Page}
% 介绍下当下热门的ordinal引爆了比特币市场， 引用下数据分析，https://dune.com/queries/2013645/3334054
% 介绍Types of Data Inscribed
% problem/challenge
% our approach % 突出文章创新点
% contribution

\begin{figure}
    \centering
    \includegraphics[scale=1.0,width=\linewidth]{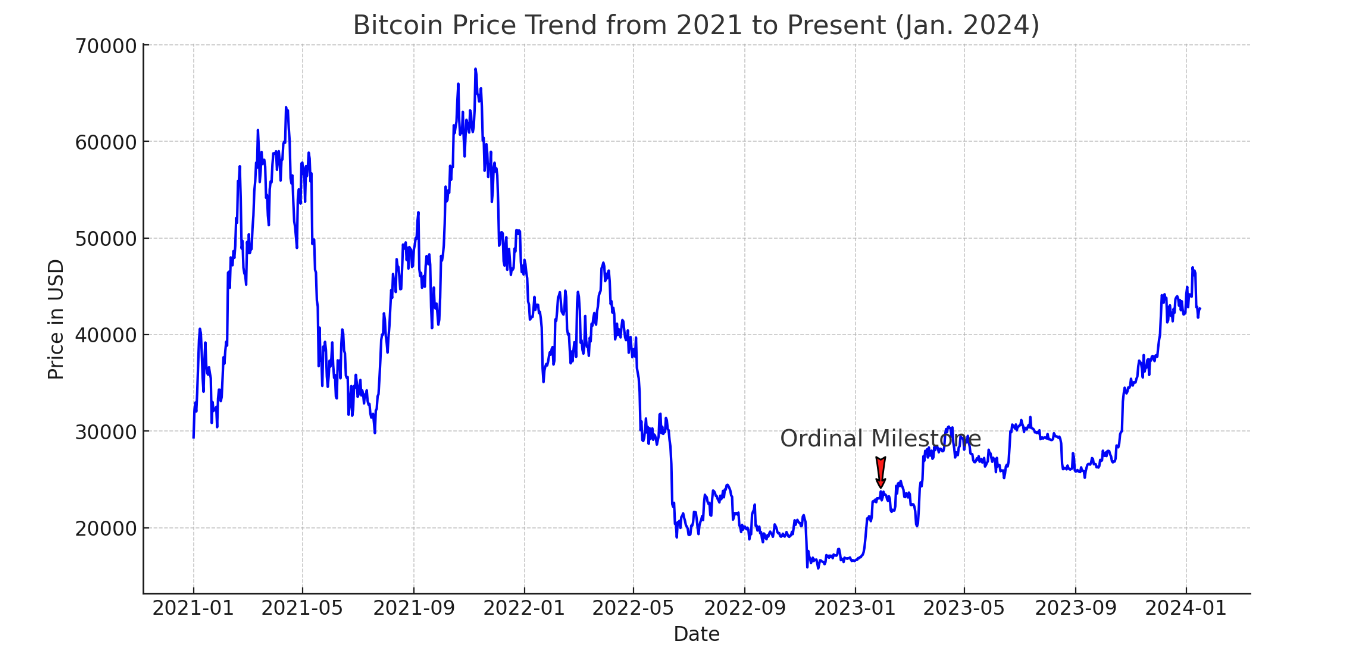}
    \caption{Bitcoin Price Trend from 2021 to Present (Jan. 2024)}
    \label{bitcoin_price_trend}
\end{figure}

When people mention blockchain, the first thing that comes to mind is Bitcoin \cite{nakamoto2008bitcoin}. Despite the past decade (an era also known as Web3~\cite{wang2022exploring}) witnessing flourishing innovation in numerous public blockchains such as Ethereum \cite{buterin2013ethereum}, BSC \cite{binancesmartchain}, and Solana \cite{yakovenko2018solana} in terms of decentralization, scalability, security, and privacy, it is undeniable that Bitcoin remains the largest and most valuable cryptocurrency asset in the world \cite{coindesk2024bitcoin}. The price trend (Fig.\ref{bitcoin_price_trend}) of Bitcoin from 2021 to the present shows that its price has experienced multiple significant rises and falls. Since reaching its historical peak of about \$69,000 in October 2021, the price of Bitcoin has been on a continuous decline. It was not until 2023 that Bitcoin welcomed a new round of ``bull market". Behind this, in addition to market sentiment and economic environment factors, the technological development of Bitcoin is the key driver igniting this round of market enthusiasm.

\begin{figure}[ht]
    \centering
    \includegraphics[scale=1.0,width=\linewidth]{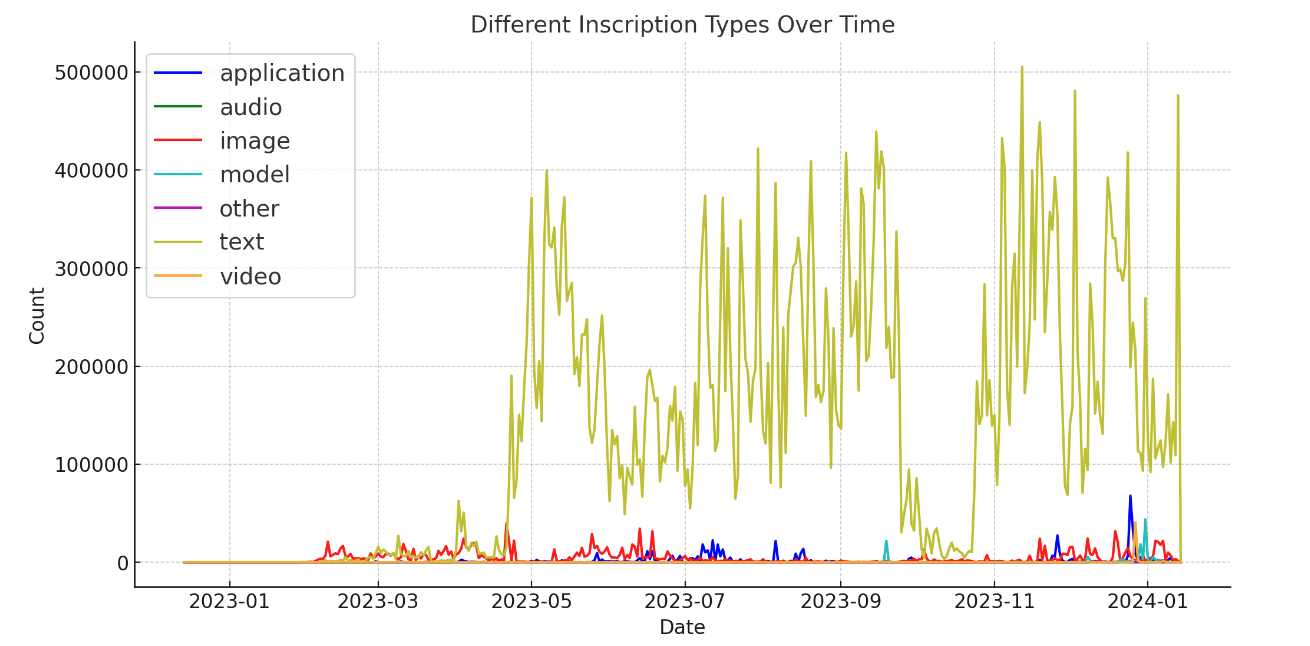}
    \caption{Different Inscription Types Over Time}
    % \qw{Legend/title can be larger}}
    \label{inscription_type}
\end{figure}

The Ordinal protocol \cite{ordinalprotocol}, crafted by Casey Rodarmor and launched on the Bitcoin mainnet on January 20, 2023, has opened new avenues for users to innovate within the Bitcoin blockchain. This is achieved through the creation of ``inscriptions", which function similarly to non-fungible tokens (NFTs) \cite{wang2021non} found on other blockchains. Specifically, it is realized by adding non-transactional data (typically in JSON style) to Bitcoin transactions. These inscriptions can include a variety of data types, ranging from simple text to complex images or code. Following its launch, Bitcoin inscriptions quickly garnered significant attention in the cryptocurrency market. As of now, there have been over 55 million \cite{glassnode2024btcinscriptions} Bitcoin inscriptions created, with text-based data types comprising over 95\% of these inscriptions (Fig.\ref{inscription_type}). Bitcoin inscriptions enhance the capabilities of Bitcoin, pushing the boundaries of what was once deemed impossible.

Previous studies have explored various aspects surrounding this new hype. Razi et al. \cite{razi2023non} present a thorough overview of NFT (including Bitcoin NFT) and survey the current landscape of NFT applications across various domains. Wang et al. \cite{wang2023understanding} delve into the surge of interest and activity surrounding BRC-20 tokens and critically examine the narratives of their hope and hype based on market sentiment. Louis \cite{bertucci2023bitcoin} delves into the factors influencing transaction fees in the context of Bitcoin Ordinals and assesses their overall impact on the Bitcoin network. Yu et al. \cite{yu2023bridging} present a detailed design of a lightweight bridge to facilitate communication and asset operation from the Bitcoin network to Ethereum blockchains. Kiraz et al. \cite{kiraz2023nft} introduce a novel mechanism for conducting NFT transactions on the Bitcoin blockchain. They propose an off-chain receipts method that allows for the certification of authenticity and ownership transfer of digital assets.

% \qw{XX et al.;  XX et al., XX et al., XX et al.,\qw{fill here (this paragraph is to include existing related literature, make the context more fluent)}.} 

While these works have contributed elegant insights, our research distinguishes itself by offering a distinct perspective. We focus on providing a clear and comprehensive interpretation of the most fundamental aspects of inscriptions, delving deeper than the surface-level usage or construction discussed in prior works. We summarise our key contributions as follows:

\begin{itemize}
    \item We provide the first technical report dedicated to the topic of Bitcoin inscriptions (Sec.\ref{sec-preliminary}), covering its necessary preliminary technologies, including SegWit, Schnorr Signatures, Taproot, Tapscript, and Ordinal Theory.

    \item We present Bitcoin inscriptions' operational mechanisms (Sec.\ref{sec-inscription}), detailed by a historical overview, an examination of the working principles, and a comparative analysis with conventional NFTs and the potential use case built on Bitcoin inscriptions highlighting their versatility across different domains, such as BRC-20, digital arts, and gaming assets.
    
    \item We discuss the challenges and opportunities from a forward-looking perspective (Sec.\ref{sec-chanlledge}), identifying areas for future research and industry development. In particular, we emphasize the promising prospects of Bitcoin Layer2 and derivative protocols related to inscriptions, drawing parallels with their success in Ethereum ecosystems.
    
\end{itemize}

%----------------------------------------------------
% Sec2 主要是介绍下涉及到的技术
\section{Entree: Preliminaries and Background}
\label{sec-preliminary}

In this section, we provide a detailed exploration of technological underpinnings, including Segregated Witness, Schnorr Signature, Taproot, Tapscript, and Ordinal Theory within the Bitcoin network.

\subsection{Segregated Witness}
The Bitcoin network consistently verifies a new block approximately every 10 to 15 minutes, with each block encompassing a specific number of transactions \cite{nakamoto2008bitcoin}. Consequently, the size of these blocks directly influences the number of transactions that can be confirmed within each block. Segregated Witness, commonly known as SegWit, represents one of the key protocol upgrades \cite{segwit2017bip} addressing scalability and transaction malleability issues.

Before SegWit, Bitcoin transactions included a component called the \textit{signature} data within the transaction structure. This data, crucial for the verification of transactions, contributed to two primary issues: scalability and transaction malleability. Specifically, each Bitcoin block has a size limit (originally set to 1MB), which constrains the number of transactions that can be processed in each block. The inclusion of signature data within transactions consumed significant space, limiting the overall transaction throughput of the network. In addition, the transaction ID (txid) was generated by hashing the entire transaction data, including the signature. Since signatures could be altered without invalidating the transaction, the txid could also be changed, leading to potential issues in tracking and confirming transactions.

\begin{figure}[ht]
    \centering
    \includegraphics[scale=1.0,width=\linewidth]{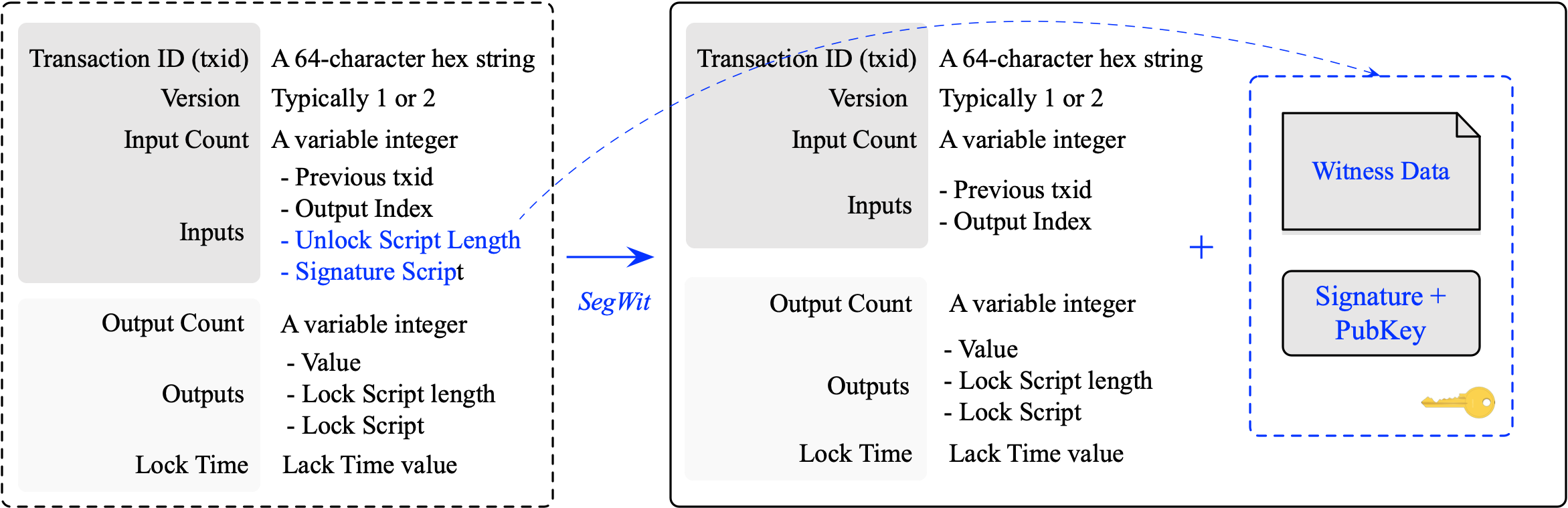}
    \caption{Bitcoin Transaction Data Structure (SegWit)}
    \label{fig:segwit}
\end{figure}

SegWit addresses these issues by changing the way transaction data is structured and stored in the blocks (cf. Fig.\ref{fig:segwit}). Firstly, A SegWit transaction consists of two main components, the original transaction structure (without the signature) and a separate \textit{witness} section containing the signatures and scripts. It is worth noting that the signature data is separated from the main transaction data. The witness information is still transmitted and stored in the blockchain but is no longer a part of the transaction's txid calculation. The txid is now calculated without including the witness data, effectively resolving the transaction malleability issue. This change means that the txid remains constant even if the signature data is altered. Secondly, SegWit introduces a new concept called \textit{block weight}, which is a blend of the block's size with and without the witness data. The maximum block weight is set to 4MB, while the size of the non-witness data is still capped at 1MB \cite{singh2020public}. This effectively allows for more transactions to be included in each block, improving the network's capacity.

We present the advantages provided by SegWit for Bitcoin inscription.

\begin{itemize}
    \item \textit{Increased inscription transactions:} SegWit increases the number of transactions that can fit into a block by changing how data is counted towards the block size limit. This indirectly benefits Bitcoin inscriptions by allowing more space for these types of transactions, which can sometimes be data-heavy.

    \item \textit{Enhanced transaction efficiency:} By optimizing the space within each block, SegWit makes the Bitcoin network more efficient. This efficiency can be beneficial for Bitcoin inscriptions, as it potentially leads to faster confirmation times and lower fees.

    \item \textit{Enhanced transaction security:} For Bitcoin inscriptions, which may rely on unaltered transaction IDs for their operation, SegWit ensures that the modification of the signature part does not affect the transaction ID. It is crucial for maintaining the integrity of inscription transaction references.
\end{itemize}
    
%-----------------------------------------------
% reference:

% BIP141: https://github.com/bitcoin/bips/blob/master/bip-0141.mediawiki

% 1. Public blockchains scalability: An examination of sharding and segregated witness - A Singh, RM Parizi, M Han, A Dehghantanha (2020)
% 2. Post Quantum Blockchain with Segregation Witness - B Li, F Wu (2021)
% 3. Analysis of segregated witness implementation for increasing efficiency and security of the Bitcoin cryptocurrency - M Kedziora, D Pieprzka, I Jozwiak, Y Liu (2023)
% 4. What drives Bitcoin fees? Using SegWit to assess Bitcoin's long-run sustainability - C Brown, J Chiu, TV Koeppl (2021)

\subsection{Schnorr Signature}
The Schnorr signature scheme was proposed by Bitcoin core developer Pieter Wuille via Bitcoin Improvement Proposal (BIP)-340 \cite{bip340} in January 2020. The proposal includes the Taproot/Schnorr soft fork upgrade \cite{taprootschnorr2020} to replace the Elliptic Curve Digital Signature Algorithm (ECDSA) \cite{johnson2001elliptic} employed in Bitcoin's digital signature mechanism.

The Schnorr signature consists of three main algorithms: \textit{key generation}, \textit{signing}, and \textit{verification}~\cite{neven2009hash}. Key generation establishes a secure private-public key pair. The signing algorithm then creates a unique signature for each message, combining a random nonce, a hash function, and the private key. The verification algorithm checks the signature's validity against the message and public key. 

\textit{Key generation.}
Key generation (Algorithm~\ref{alg:key_generation}) initiates the Schnorr signature process. It involves generating a private key $d$ and a public key $Q$. The private key is randomly selected from the set $\mathbb{Z}_n^*$, representing the group of integers modulo $n$ (excluding zero), where $n$ is the order of the elliptic curve. The public key $Q$ is computed as the product of the private key $d$ and the generator point $G$ of the elliptic curve. The generator point $G$ is a pre-defined point on the elliptic curve, known to all parties in the cryptographic system. The output of this algorithm is the pair $(d, Q)$.

\begin{algorithm}
    \caption{Key generation}\label{alg:key_generation}
    \begin{algorithmic}[1]
        \REQUIRE Generator point $G$, order of the curve $n$
        \ENSURE Private key $d$, Public key $Q$
        \STATE $d \gets$ randomly choose from $\mathbb{Z}_n^*$
        \STATE $Q \gets d \cdot G$
        \RETURN $(d, Q)$
    \end{algorithmic}
\end{algorithm}

\textit{Signing procedure.}
Signing (Algorithm~\ref{alg:sign}) is responsible for creating a digital signature for a given message. It starts by selecting a random nonce $k$ from $\mathbb{Z}_n^*$. This nonce is a random number that ensures the uniqueness of each signature, even for repeated signings of the same message. The algorithm computes a point $R$ on the elliptic curve by multiplying $k$ with the generator point $G$. Subsequently, it calculates a hash value $e$ using a cryptographic hash function, taking as input the concatenation of $R$, the public key $Q$, and the message $m$. The signature component $s$ is then computed as $(k - e \cdot d) \mod n$, where $d$ is the private key. The resulting signature for the message $m$ is the pair $(R, s)$.

\begin{algorithm}
    \caption{Signing}
    \begin{algorithmic}[1]
        \REQUIRE Private key $d$, message $m$
        \ENSURE Signature $(R, s)$
        \STATE $k \gets$ randomly choose from $\mathbb{Z}_n^*$
        \STATE $R \gets k \cdot G$
        \STATE $e \gets \text{Hash}(R \parallel Q \parallel m)$
        \STATE $s \gets (k - e \cdot d) \mod n$
        \RETURN $(R, s)$
    \end{algorithmic}
    \label{alg:sign}
\end{algorithm}

\textit{Signature validation.}
Verification (Algorithm~\ref{alg:verification}) determines the validity of a given signature for a message. It requires the public key $Q$, the message $m$, and the signature $(R, s)$ as inputs. The algorithm first computes the hash value $e$ similarly to the signing algorithm. It then calculates a verification point $V$ on the elliptic curve, which is the sum of $s \cdot G$ and $e \cdot Q$. The signature is deemed valid if and only if this verification point $V$ equals the point $R$ in the signature. If $V = R$, the algorithm returns \textit{True}, indicating the signature is valid. Otherwise, it returns \textit{False}, indicating the signature is invalid.

\begin{algorithm}
    \caption{Verification}
    \begin{algorithmic}[1]
        \REQUIRE Public key $Q$, message $m$, signature $(R, s)$
        \ENSURE Validation result
        \STATE $e \gets \text{Hash}(R \parallel Q \parallel m)$
        \STATE $V \gets (s \cdot G) + (e \cdot Q)$
        \IF{$V = R$}
            \RETURN $\text{True}$ \COMMENT{\textcolor{black}{The signature is valid}}
        \ELSE
            \RETURN $\text{False}$ \COMMENT{\textcolor{black}{The signature is invalid}}
        \ENDIF
    \end{algorithmic}
    \label{alg:verification}
\end{algorithm}

Employing Schnorr signatures is particularly important to Bitcoin inscription technology for several reasons that align well with the cryptocurrency's goals of security, efficiency, and privacy. Here are the key reasons.

\begin{itemize}
    \item \textit{Efficient data embedding:} Schnorr signatures, known for their efficiency in terms of size, enable more data to be embedded within a transaction while minimizing the space it occupies on the blockchain \cite{jain2023sok}. This efficiency is crucial for inscription technology, which involves embedding additional data (like digital artifacts) on the blockchain. The compact nature of Schnorr signatures allows for more inscriptions without significantly increasing the size of the blockchain.

    \item \textit{Enhanced privacy and security:} Inscription can potentially expose more transaction details. Schnorr signatures help mitigate privacy concerns by making multi-signature transactions indistinguishable from single-signature ones. This feature is vital in maintaining user privacy, especially when inscriptions are used in transactions involving multiple parties.
    
    \item \textit{Scalability and throughput:} The reduced size and increased efficiency of Schnorr signatures directly contribute to Bitcoin's scalability. By allowing more transactions (and thus more inscriptions) to be included in each block, Schnorr signatures can improve the overall throughput of the Bitcoin network. This is particularly important as the volume of transactions, including those with inscriptions, continues to grow.
    
    \item \textit{Flexibility for complex transactions:} Schnorr signatures offer greater flexibility for complex transactions, which is beneficial for advanced inscription use cases. They enable sophisticated scripting possibilities, which can be used to create more intricate types of inscriptions and digital artifacts on Bitcoin.

\end{itemize}

% BIP 340: https://lists.linuxfoundation.org/pipermail/bitcoin-dev/2018-July/016203.html

% Casas, P., Romiti, M., Holzer, P., Mariem, S. B. (2021). Where is the light (ning) in the taproot dawn? Unveiling the bitcoin lightning (IP) network. IEEE 10th International Conference.
% Jain, A., Pilli, E. S. (2023). SoK: Digital Signatures and Taproot Transactions in Bitcoin. International Conference on Information Systems Security and Privacy.
% Kleinwort, F., Posdorfer, W. (2023). Analyzing the Effect of Taproot on Bitcoin Deanonymization. IEEE 43rd International Conference.

\subsection{Taproot and Tapscript}
BIP-341 \cite{bip341}, commonly known as Taproot, is an enhancement to BIP-340. The core idea is to combine the strengths of \textit{Merkelized abstract syntax trees} MAST \cite{bip114} and \textit{Schnorr signatures} by committing a single Schnorr public key in the output that can represent both a single public key spend and a complex script spend. It introduces a new Taproot output (SegWit version 1) that includes a signature, a control block, and a script path. Moreover, it also specifies the rules for spending the Taproot output, which can be either a key-path spend (using a single signature) or a script-path spend (using one of the scripts committed to in the MAST structure). It achieves privacy and efficiency by allowing users to mask complex smart contracts as standard single-signature transactions.

Tapscript proposed in BIP-342 \cite{bip342} introduces improvements to the Bitcoin scripting language, particularly focusing on the integration of Schnorr signatures and used for Taproot script-path spends. The primary motivation behind BIP-342 is to address certain limitations within the existing Bitcoin scripting system, especially in terms of compatibility with the semantics of certain opcodes. BIP-342 modifies the signature opcodes \texttt{OP\_CHECKSIG} and \texttt{OP\_CHECKSIGVERIFY} to verify Schnorr signatures as specified in BIP-340. It presents a new opcode, \texttt{OP\_CHECKSIGADD}, which facilitates the establishment of multi-signature policies that can be verified in batches. This enhancement significantly boosts the efficiency and scalability of transactions involving multiple signatures. 

\begin{figure}[ht]
    \centering
    \includegraphics[scale=1.0,width=\linewidth]{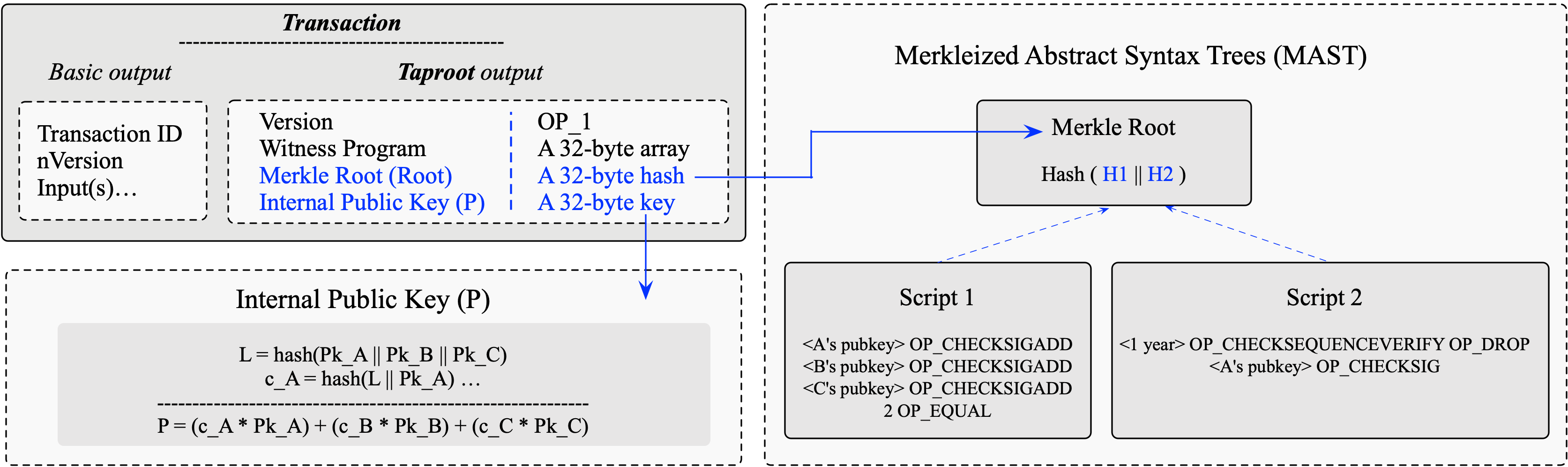}
    \caption{Taproot and Tapscript}
    \label{fig:taproot}
\end{figure}

In comparison, BIP-341's Taproot allows for the compression of data in inscription transactions, reducing their size and improving scalability. BIP-342 enables batch verification of signatures, which is crucial for multi-signature transactions often used in inscriptions.  To further illustrate the working principle of BIP-341 and BIP-342, we consider a situation where a Bitcoin address is controlled by three parties: A, B, and C. We created a Taproot output address that allows for flexible spending conditions using MAST and Tapscript (shown in Fig.\ref{fig:taproot}). The spending conditions are as follows:
\textit{1. Any two of the three parties (A, B, C) can jointly spend the funds.}
\textit{2. If the funds are not moved for a year, party A can unilaterally spend them (a time-lock condition).} Initially, the parties aggregate their individual public keys (Pk\_A, Pk\_B, Pk\_C) to form a single Schnorr Internal public key P. This aggregate key, alongside the Merkle root derived from two hashed scripts (Script1 adds up the valid signatures and ensures that at least two signatures are provided; Script2 uses \texttt{OP\_CHECKSEQUENCEVERIFY} to enforce the time-lock, ensuring the script can only be executed after a year. Then it checks the signature from A), and forms the Taproot output. When it's time to spend, the parties can opt for a key-path spend using P for a private transaction, or a script-path spend, revealing the chosen script and its corresponding Merkle proof to fulfill specific conditions. 

% \qw{Add one more preliminary? Taproot, BIP-340(done)/341/342}

\subsection{Ordinal Theory}
Bitcoin Ordinal Theory \cite{ordinaltheory2024} introduces the uniqueness within the Bitcoin ecosystem. It revolves around the idea of assigning distinct identities to individual satoshis (sats), the smallest unit of Bitcoin (100 millionths of a Titcoin)~\cite{nakamoto2008bitcoin}, allowing for their precise tracking and utilization in various applications.

\smallskip
\noindent\textbf{Numbering scheme for satoshis.}
Satoshis are numbered in the order they are mined, and this numbering is maintained as they are transferred from transaction inputs to outputs, adhering to a first-in-first-out principle. Ordinal numbers are essentially a numbering scheme for satoshis, allowing each satoshi to be tracked and transferred as an individual entity.

Ordinal numbers can be represented in several distinct formats \cite{ordinaltheory2024}. Taking Inscription \hlhref{https://ordinals.com/sat/1938930000000000}{\#34,595,802} as a case in point. %https://ordinals.com/sat/1938930000000000 

\begin{itemize}
    \item \textit{Integer notation:} The ordinal number assigned according to the order in which the satoshi was mined. E.g., 1938930000000000.
    \item \textit{Decimal notation:} Combines the block height at which the satoshi was mined with the offset of the satoshi within the block. E.g., 792288.0.
    \item \textit{Degree notation:} A unique representation that makes the rarity of a satoshi easy to see at a glance. E.g., \( 0^\circ 162288' 0'' 0''' \).
    \item \textit{Percentile notation:} The satoshi's position in Bitcoin's supply is expressed as a percentage. E.g., 92.33000010156304\%.
    \item \textit{Name:} An encoding of the ordinal number using characters a through z. E.g., acqgzfkezav.
\end{itemize}

\smallskip
\noindent\textbf{Rarity system.}
Ordinal Theory introduces a system of rarity based on periodic events in Bitcoin, such as blocks, difficulty adjustments, halvings, and cycles. This system categorizes satoshis into different rarity levels like common, uncommon, rare, epic, legendary, and mythic \cite{ordinaltheory2024}.

Bitcoin experiences regular events that vary in frequency, creating a natural framework for categorizing rarity. These events include:
\begin{itemize}
    \item \textit{Blocks:} New blocks are mined about every 10 minutes.
    \item \textit{Difficulty adjustments:} Occurring every 2016 blocks or roughly every two weeks. The network adjusts the difficulty level required for block acceptance in response to hash rate fluctuations.
    \item \textit{Halvings:} Approximately every four years, or every 210,000 blocks, the number of new satoshis generated per block is halved.
    \item \textit{Cycles:} A special event, known as a conjunction, happens every six halvings, approximately every 24 years \footnote{The LCM is the smallest number that is a multiple of 2016 and 210,000.:
        \[
        \text{1260,000 = LCM}(2016, 210,000)
        \]
    },
    where halvings and difficulty adjustments align. The period between these conjunctions is termed a cycle, with the next one anticipated around 2032.
\end{itemize}

Based on these events, satoshis can be classified into different levels of rarity:

\begin{itemize}
    \item \textit{Common:} Any satoshi that isn't the first in its block.
    \item \textit{Uncommon:} The first satoshi in each block.
    \item \textit{Rare:} The first satoshi in each difficulty adjustment period.
    \item \textit{Epic:} The first satoshi in each halving epoch.
    \item \textit{Legendary:} The first satoshi in each cycle.
    \item \textit{Mythic:} The very first satoshi from the genesis block.
\end{itemize}

For example, Inscription \hlhref{https://ordinals.com/sat/1938930000000000 }{\#34,595,802} is rare. We can use degree notation to highlight the rarity of this satoshi straightforwardly.
\[
1^\circ 1' 0'' 0''' \quad \\
\]

\begin{tabular}{cl} 
    \hline
    $1^\circ$ & Second \textbf{cycle} \\
    $1'$ & Not the first block in \textbf{halving epoch} \\
    $0''$ & First block in difficulty \textbf{adjustment period} \\
    $0'''$ & First sat in \textbf{block} \\
\end{tabular}

\subsubsection{Impact of Ordinal Theory on Bitcoin inscriptions:}

Ordinal Theory allows for the creation of unique, non-fungible assets directly on the Bitcoin blockchain. By assigning distinct identities to individual satoshis, it becomes possible to attach specific data or digital artifacts to these satoshis, effectively turning them into unique digital assets. Additionally, with Ordinal Theory, each satoshi can be tracked through its transaction history. This level of traceability is crucial for establishing the provenance and ownership history of inscribed assets.

%-----------------------------------------------------
%Sec 3 着重介绍什么是比特币铭文（commonly known as NFT）；比特币铭文是怎么实现的；与其他nft的特性对比；

\section{Staple-I: Understanding Bitcoin Inscriptions}\label{sec-inscription}
This section offers a comprehensive understanding of Bitcoin inscriptions, illustrating the history of Bitcoin-related NFTs, technical working principles, a comparative analysis distinguishing Bitcoin inscriptions from NFTs on platforms like Ethereum, and the exploration of use cases.

\subsection{Historical Overview} 

Bitcoin's creation in 2009 laid the groundwork for blockchain technology, but initially, it was focused on cryptocurrency rather than NFTs. Namecoin in 2011 \cite{kalodner2015empirical} was a pioneering project that forked from Bitcoin. It focused on using blockchain technology for decentralized domain name registration, representing an early exploration of blockchain's potential beyond currency. Its limitations included limited adoption and being overshadowed by Bitcoin's growth.

One of the earliest concepts related to NFTs on the Bitcoin blockchain was ``Colored Coins" \cite{rosenfeld2012overview} in 2012. Colored Coins implemented a method known as ``metadata injection" to incorporate additional metadata. This process utilized the scriptSig field within a transaction's input script. However, this approach was inefficient and faced limitations because it was not designed for storing extensive metadata, leading to scalability issues.

In 2014, Counterparty \cite{counterparty2014} emerged as a significant development, building on the Bitcoin blockchain to enable the creation of unique tokens and digital assets. This platform expanded the possibilities for asset representation and transfer on Bitcoin. However, it was limited by Bitcoin's transaction speed and fee structure. The reliance on Bitcoin's blockchain meant that Counterparty transactions were subject to the same congestion and high fees during peak times.

Spells of Genesis \cite{spellsofgenesis2015} proposed in 2015, one of the first games to integrate blockchain technology, utilized Counterparty to create in-game assets as tradable tokens. While it represented a novel blend of gaming and blockchain, the game's dependence on the Counterparty platform inherited its limitations.

In 2016, Rare Pepe \cite{rarepepes2016} brought a unique cultural twist to blockchain tokens. Using Counterparty, it allowed the creation and trading of digital assets based on the popular Pepe the Frog meme, showcasing the diverse potential of blockchain assets. However, it was constrained by the niche appeal of its meme-based assets and the technical limitations of the Counterparty platform. The project's success was more within the crypto community than in broader markets.

More recently, the concept of Bitcoin NFTs has seen a resurgence with the introduction of Bitcoin Ordinals \cite{ordinaltheory2024}. This approach involves assigning unique ordinal numbers to individual satoshis, enabling them to be used as distinct, trackable assets akin to NFTs. This has opened up new possibilities for creating and trading digital artifacts directly on the Bitcoin blockchain. Despite this, Bitcoin NFTs are still in their nascent stages. Challenges include potential network congestion due to increased data from inscriptions and the permanence of content issues (more discussions refer to Sec.\ref{sec-chanlledge}).

\begin{table}[!hbt]
\caption{Historical Overview of Bitcoin Inscriptions}
\label{table:bitcoin-nfts}
\centering
\rowcolors{2}{gray!15}{white} % Alternate row colors
\begin{tabular}{>{\bfseries}l l p{4cm} p{4.5cm}}
\toprule
\rowcolor{gray!25} % Header row color
\textbf{Year} & \textbf{Milestone} & \textbf{Description} & \textbf{Limitations} \\
\midrule
2011 & Namecoin & Early fork of Bitcoin for decentralized domain name registration. & Limited adoption and overshadowed by Bitcoin's growth. \\
2012 & Colored Coins & Early asset representation method on Bitcoin. & Inefficient, capacity constraints. \\
2014 & Counterparty & Platform for Bitcoin-based unique tokens. & Bitcoin's transaction speed and fee limitations. \\
2015 & Spells of Genesis & Blockchain game using Counterparty for assets. & Limited mainstream adoption, platform constraints. \\
2016 & Rare Pepe & Digital assets based on Pepe meme via Counterparty. & Niche appeal, technical limitations. \\
2023 & Bitcoin Ordinals & Advancements for digital artifacts on Bitcoin. & Network congestion, content permanence issues. \\
\bottomrule
\end{tabular}

\end{table}

\subsection{Working Principle}
\begin{figure}[!ht]
    \centering
    \includegraphics[scale=1.0,width=\linewidth]{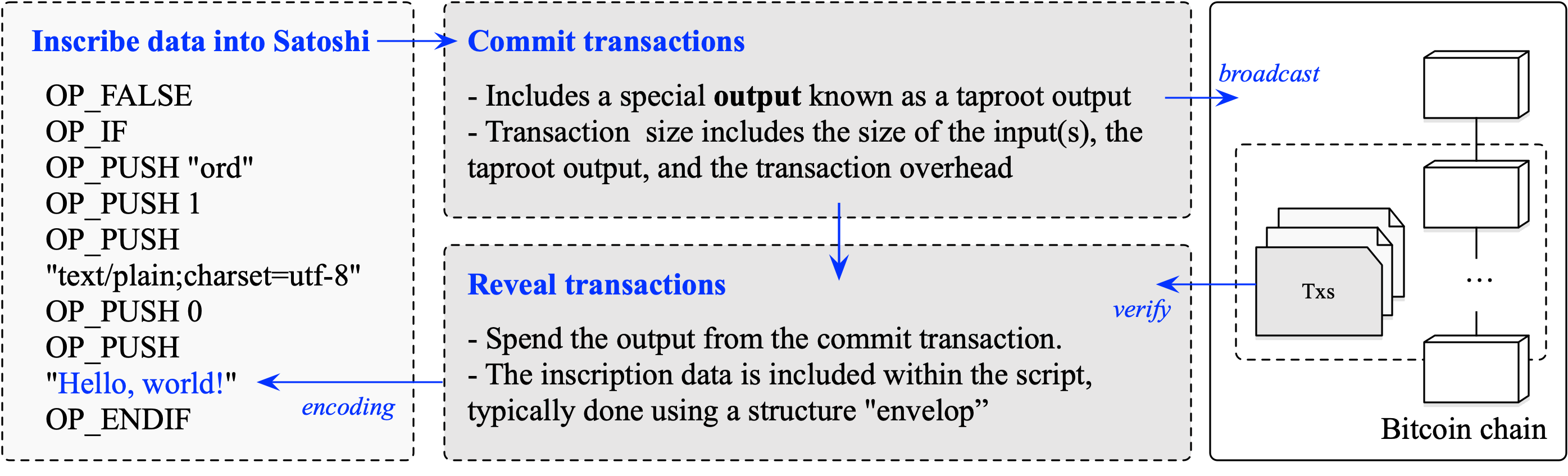}
    \caption{Working Principle of Bitcoin Inscriptions}
    \label{inscription_process}
\end{figure}

The following diagram (Fig.\ref{inscription_process}) illustrates the process of inscribing data onto the Bitcoin blockchain and associating it with a specific ordinal (satoshi).

The working principle is as follows: Firstly, the data intended for inscription is first prepared in a suitable format. This often involves encoding the data into a byte string that can be embedded in a Bitcoin transaction. The data is typically accompanied by a MIME type \cite{shafranovich2005common} that specifies the nature and format of the content (e.g., image/jpeg, text/plain), ensuring compatibility with web standards. This alignment with web standards means that inscription content can be retrieved and rendered by standard web browsers, similar to how they handle regular web content served from a server.

Then, inscription content is encapsulated within a structure known as an ``envelope'', which uses Bitcoin taproot script-path spend script \cite{ordinaltheory2024} opcodes \texttt{OP\_FALSE}, \texttt{OP\_IF}, …, \texttt{OP\_ENDIF}. Taproot scripts benefit from the witness discount \cite{taprootschnorr2020} introduced by SegWit, which reduces the size and cost of storing witness data compared to other transaction data.

Creating the inscription transaction has two phases. In the first phase (called \textit{commit transaction}), a taproot output is created, which commits to a script containing the inscription content. This script is constructed in such a way that it cryptographically references the data without revealing it. The transaction is then broadcast and included in a block on the blockchain. In the second phase, a subsequent transaction (\textit{reveal transaction}) is made to spend the output from the commit phase. This spending transaction includes the actual inscription data within its script. When this transaction is confirmed, the inscription data becomes permanently recorded on the blockchain.

Finally, the unique ordinal number of the selected satoshi is tracked through these transactions. As ordinals follow a first-in-first-out principle, such ordinal number is effectively transferred from the input satoshi to the output satoshi (the commit transaction) and then to the inscribed satoshi (the reveal transaction).

\subsection{Differences Between Bitcoin Inscriptions and Other NFTs}

Bitcoin inscriptions and NFTs on other blockchains (majorly, EVM-compatible platforms) represent two distinct approaches. This comparative analysis (summarised in Table \ref{table:bitcoin-nfts-comparison}) aims to discuss the fundamental differences between these two paradigms, highlighting their unique characteristics and implications.

\begin{table}[h]
\caption{Bitcoin Inscriptions vs. NFTs on Other Blockchains}
\label{table:bitcoin-nfts-comparison}
\centering
\rowcolors{2}{gray!15}{white} % Alternate row colors
\begin{tabular}{>{\bfseries}l p{4cm} p{4cm}}
\toprule
\rowcolor{gray!25} % Header row color
\textbf{Aspect} & \textbf{Bitcoin Inscriptions} & \textbf{Other NFTs} \\
\midrule
Description Method & Digital Asset, Inscription, Oridinal & NFT \\
Protocol Form & Ordinal Protocol & ERC-721, ERC-1155, etc \\
Storage Method & Entirely stored on-chain & Stored on IPFS or Arweave, not 100\% on-chain \\ 
Immutability & Inherent & Variable \\
Minting & Currently not possible without a node, only possible via third-party designed services & Mostly can directly interact with the webpage \\
Trading Method & NFT Marketplace & NFT Marketplace \\
Scarcity  & Limited by Bitcoin usage & Potentially less scarce \\
Royalty Models & None & Common, with challenges \\
Integration & Difficult due to Bitcoin's scripting limitations & Easier due to programmable smart contracts \\ 
Energy Consumption & High due to Proof of Work consensus & Depends on the consensus algorithm of the platform \\
Advantages & Scarcity, characteristics of luxury goods & Mainstream NFT mode, high user base \\ 
Disadvantages & 1. Block speed is slow, and not suitable for bulk minting. 2. Difficulties in minting and trading. 3. Wallet entry is complex & No special gimmicks or fame, easily overlooked \\

\bottomrule
\end{tabular}

\end{table}

% \qw{could merge the following unmentioned items into Table2}
% \begin{longtable}{p{0.25\textwidth}|p{0.35\textwidth}|p{0.35\textwidth}}
% \toprule

% \textbf{Description} & \textbf{Bitcoin NFT} & \textbf{Other NFT} \\ 
% \midrule
% Protocol Form & Ordinal & ERC-721, ERC-1155, SPL, etc \\ [1ex] 
% \midrule
% General Description Method & Inscription & NFT \\ [1ex] 
% \midrule
% Storage Method & Entirely stored on-chain & Stored on IPFS or Arweave, not 100\% on-chain \\ [1ex] 
% \midrule
% Code Modification & Not allowed & Depends on project's contract code \\ [1ex] 
% \midrule
% Minting & Currently not possible without a node, only possible via third-party designed services & Mostly can directly interact with the webpage \\ [1ex] 
% \midrule
% Trading Method & NFT Marketplace & NFT Marketplace \\ [1ex] 
% \midrule
% Integration with Other Platforms & Difficult due to Bitcoin's scripting limitations & Easier due to programmable smart contracts \\ [1ex] 
% \midrule
% Energy Consumption & High due to Proof of Work consensus & Depends on the consensus algorithm of the platform \\ [1ex] 
% \midrule
% Advantages & Scarcity, characteristics of luxury goods & Mainstream NFT mode, high user base \\ [1ex] 
% \midrule
% Disadvantages & 1. Block speed is slow, and not suitable for bulk minting. 2. Difficulties in minting and trading. 3. Wallet entry is complex & No special gimmicks or fame, easily overlooked \\ [1ex] 

% \bottomrule

% \end{longtable}

\textit{Protocol and description method.} Bitcoin inscriptions operate on the Ordinal protocol, embedding data directly into individual satoshis, and are described as digital assets, ordinals, or inscriptions. In contrast, NFTs on platforms like Ethereum utilize standards such as ERC-721 \cite{erc721standard} or ERC-1155 \cite{erc1155standard}, and are commonly referred to as NFTs. This fundamental difference in protocol form shapes the inherent properties and possibilities of each asset type.

\textit{Storage and immutability.} Bitcoin inscriptions are entirely stored on-chain, ensuring complete immutability and permanence. Once an inscription is embedded in the Bitcoin blockchain, it becomes unalterable. This method ensures the durability of the inscribed content, with the cost of inscriptions being proportional to the content's size. On the other hand, NFTs on other blockchains often rely on off-chain storage solutions like IPFS \cite{benet2014ipfs} and Arweave \cite{arweaveWhitepaper}, introducing potential challenges in content availability and permanence. The degree of immutability for these NFTs can vary, with some being alterable or deletable by the contract owner. This variability necessitates a thorough audit of the contract code to ascertain the immutability of an NFT. The auditing process can be complex and technically demanding.

\textit{Minting and trading.} Minting Bitcoin inscriptions currently requires a node or third-party services, and the trading of these inscriptions occurs through NFT marketplaces. Other NFTs can often be minted directly through web interfaces and are also traded on NFT marketplaces. However, the integration of Bitcoin inscriptions with other platforms is challenging due to Bitcoin's scripting limitations, while NFTs on programmable blockchains offer more seamless integration.

\textit{Scarcity.} Bitcoin inscriptions are inherently limited by the nature of the Bitcoin blockchain. Since they are directly tied to individual satoshis, their scarcity is intrinsically linked to the total supply of Bitcoin, which is capped at 21 million coins. This finite supply of satoshis imposes a natural limit on the number of possible inscriptions. On platforms like Ethereum, the creation of NFTs is often governed by smart contracts that allow for flexible minting policies. Creators can generate large quantities of NFTs with minimal cost, which can potentially lead to a saturation of the market. While some NFT projects on other blockchains impose a cap on the number of tokens, others do not, allowing for an unlimited supply. This flexibility means that the scarcity of NFTs on these platforms can vary significantly from one project to another.

\textit{Royalty models.} Bitcoin inscriptions do not inherently support on-chain royalty models. This means that when a Bitcoin inscription is transferred or sold, there is no automatic mechanism within the Bitcoin blockchain to provide a percentage of the sale back to the original creator. Many NFTs on platforms like Ethereum utilize smart contracts that can include on-chain royalty models. These contracts automatically enforce that a certain percentage of every secondary sale of the NFT is paid to the original creator. The presence of on-chain royalty models offers ongoing compensation for creators but introduces complexity and standardization challenges.

\textit{Integration and energy consumption.} Bitcoin inscriptions benefit from direct integration into the largest and most established cryptocurrency market, with immediate access to a vast and liquid market, but it is difficult to integrate into other platforms due to Bitcoin's scripting limitations. On the other hand, their energy consumption is high due to Bitcoin's Proof of Work consensus. In contrast, NFTs on other blockchains are confined to specific ecosystems with varying energy footprints based on the consensus algorithm of the platform.

\textit{Advantages and disadvantages.}
Bitcoin inscriptions offer the advantages of scarcity and the characteristics of luxury goods. However, they face challenges such as slow block speed, complexities in minting and trading, and a complex wallet setup. Other NFTs enjoy a mainstream mode and a high user base but may lack unique features or recognition, potentially leading to market saturation.

%-----------------------------------------------------
%Sec 4 (now moved into sec3)
\subsection{Use Cases}\label{sec-usecase}

\noindent\textbf{Token Standard for Fungible Tokens on Bitcoin: BRC-20.}
%domo twitter: https://twitter.com/domodata/status/1633658974686855168?s=20
On March 8, 2023, an anonymous developer named \@domodata launched the BRC-20 \cite{brc20whitepaper}. The BRC-20 token standard represents an experimental approach to creating fungible tokens using ordinal inscriptions. This standard is akin to Ethereum's ERC-20 \cite{erc20standard} but is uniquely tailored for the Bitcoin network~\cite{wang2023understanding}. Specifically, it requires users to fill in various parts according to a standardized JSON format, with the following specifications as shown in Fig.\ref{fig:brc20}:
\begin{figure}[ht]
    \centering
    \includegraphics[scale=1.0,width=\linewidth]{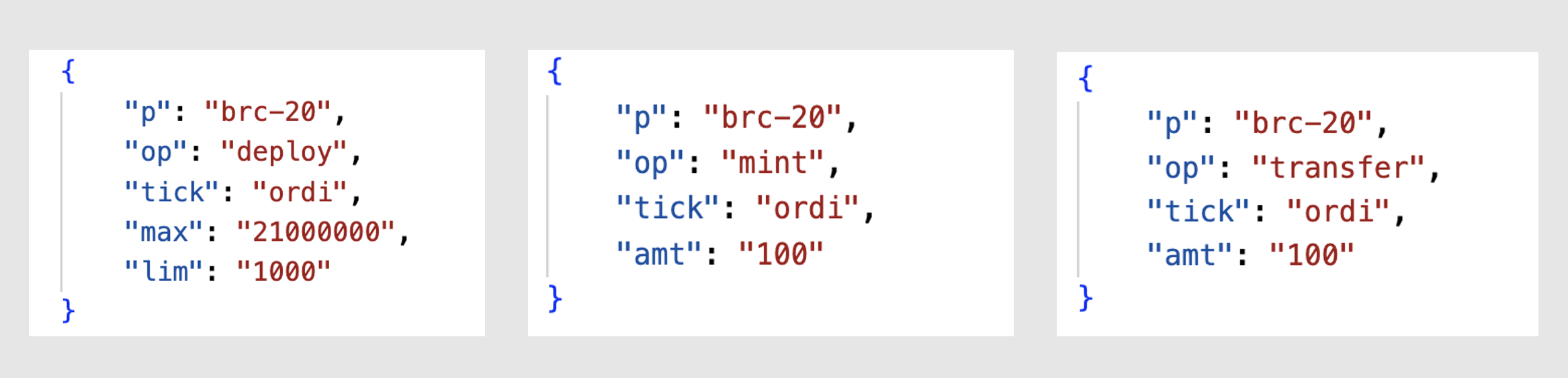}
    \caption{BRC-20 Protocol Standard}
    \label{fig:brc20}
\end{figure}

\begin{itemize}
    \item \textit{`p': protocol type.} This is a mandatory keyword that defines the operation based on the BRC-20 protocol, helping other systems to identify and process BRC-20 events.
    
    \item \textit{`op': event type.} This is a mandatory keyword that defines the type of event, whether it's \texttt{deploy}, \texttt{mint}, or \texttt{transfer}. For example, the content of `op' as `transfer' means the event type is a transfer.
    
    \item \textit{`tick': BRC-20 token identifier.} This is a mandatory keyword that defines the name of the BRC-20 token, composed of 4 letters. In this case, the content of `tick' as `ORDI' means the BRC-20 token being transferred is \$ORDI.
   
    \item \textit{`amt': the amount of BRC-20 token being transferred.} This is a mandatory keyword that defines how many BRC-20 Tokens will be transferred.
   
    \item \textit{`max': maximum supply.} This is a mandatory keyword that defines the maximum supply of the BRC-20 token.
  
    \item \textit{`lim': the maximum number of BRC-20 tokens that a single inscription can accommodate.} This is an optional keyword that defines how many BRC-20 tokens a user can obtain from minting a single inscription. If set to 1000, minting a single inscription can obtain a maximum of 1000 BRC-20 tokens.
\end{itemize}

ORDI \cite{cryptoOrdinalsPrice} is the first BRC-20 token issued on the Bitcoin network, with a total supply of 21,000,000 tokens. Unlike Ethereum's ERC-20 tokens, ORDI is not a smart contract token. It operates without the support of underlying technology, project teams, real-world project applications, or defined use cases. The valuation of ORDI is essentially driven by community consensus and the dynamics of market interest (also known as the meme coins \cite{yousaf2023connectedness}). Creating a new meme coin comes at no cost; each one faces competition from other, more 'meaningful' meme coins. For instance, another BRC-20 token, SATS \cite{cointelegraph2024ordisats}, competes in this space. Each Bitcoin can be divided into 100 million satoshis, and this BRC-20 token named SATS has set its total supply at 21 trillion, mirroring the maximum number of 21 trillion satoshis. 

\smallskip
\noindent\textbf{Digital art and collectibles.}
Bitcoin inscriptions can offer artists and collectors an innovative approach to creating and trading digital works. It leverages the capability to inscribe digital assets directly onto individual satoshis on the Bitcoin network, ensuring each piece's authenticity and permanence. A notable example of this is the creation of Ordinal Punks \cite{magiceden2023}, which draw inspiration from the iconic CryptoPunks \cite{cryptopunks2023} on the Ethereum network. These pixel-art characters, each with unique features and levels of rarity, are inscribed onto satoshis, turning them into highly desirable collectibles within the digital art sphere. Furthermore, writers and poets are finding a new medium for their work through Bitcoin inscriptions. Entire poems or short stories can be inscribed onto satoshis, creating a unique form of literary art. For instance, a poem by Ana Maria Caballero sold for 0.28 Bitcoin (equiv. \$11,430) at Sotheby’s \cite{caballero2024}, one of the world's most famous auction houses.

\smallskip
\noindent\textbf{Gaming assets.}
Leveraging the immutable nature of the Bitcoin blockchain, gaming assets inscribed as ordinals provide players with ownership of unique in-game items, characters, or even entire game worlds. The ``Pizza Ninjas" gaming project \cite{pizzaninjas2023} intertwines the thrill of gaming with the world of digital collectibles. Each Pizza Ninja character is not only a playable asset but also a piece of art. By inscribing these unique characters and narratives onto the Bitcoin blockchain, the project ensures the rarity and ownership of every in-game element.

\smallskip
\noindent\textbf{Record of messages.}
The use of Bitcoin inscriptions for creating an immutable record of messages is another feasible application. Inscribing messages can serve as proof of existence for documents or intellectual property. By embedding a hash of the document or a reference to the work in a Bitcoin transaction, creators can prove the existence of their work at a specific point in time. In some scenarios where the integrity of communication is crucial, such as in legal contexts, this way ensures that the content of the communication remains unaltered and verifiable.

%-----------------------------------------------------
%Sec 5
\section{Staple-II: Our Discussion}\label{sec-chanlledge}

\subsection{Challenges}
% \subsubsection{Technical Challenges}
% The technical challenges associated with Bitcoin inscriptions can be broadly categorized into several key areas.

\noindent\textbf{Blockchain bloat.} It refers to the increase in the size of the blockchain due to the accumulation of data over time \cite{sohan2021increasing}. In the context of Bitcoin inscriptions, blockchain bloat is exacerbated by the addition of non-financial data embedded directly into the blockchain. As the blockchain grows in size, the storage requirements for nodes also increase. In addition, new nodes or nodes catching up to the current state of the blockchain may experience slower synchronization times due to the larger amount of data that needs to be processed and verified. Moreover, larger block sizes can impact the efficiency of the network and potentially undermine the decentralized nature of Bitcoin.

\smallskip
\noindent\textbf{Limited smart contracts and compatibility.} It refers to Bitcoin's inherent design, which prioritizes security and simplicity over complex programmability. Unlike platforms like Ethereum, which are designed to support a wide range of decentralized applications through smart contracts, Bitcoin's scripting language is more restricted. This limitation impacts the blockchain's ability to support complex transactions and scalable applications, particularly those requiring intricate logic and interactions, such as DeFi protocols~\cite{jiang2023decentralized}.

\smallskip
\noindent\textbf{Security threats.}
The United States National Vulnerability Database (NVD) has identified a significant cybersecurity risk associated with the Ordinals Protocol in Bitcoin \cite{cve202350428}. This recognition highlights a significant security flaw associated with the development of the Ordinals Protocol in Bitcoin. The vulnerability allows for bypassing the datacarrier limit by disguising data as code in certain versions of Bitcoin Core and Bitcoin Knots. This exploit has been utilized by inscriptions, leading to the addition of substantial non-transactional data to the blockchain. Moreover, with the rising popularity of Bitcoin inscriptions, this ecosystem has attracted various types of scams. For example, scammers construct JSON fields for transferring inscriptions and encode them as hex for users to inscribe \cite{goplus2023inscriptionScams}. This can result in the theft of users’ inscriptions. Additionally, on trading platforms, users often encounter numerous inscriptions with the same name, making it challenging to identify authentic ones. Scammers exploit this by forging inscription series and adding invalid fields to mimic genuine ones.

% https://nvd.nist.gov/vuln/detail/CVE-2023-50428

% https://www.hstoday.us/subject-matter-areas/cybersecurity/us-agency-identified-bitcoin-inscriptions-as-cybersecurity-risk/#:~:text=The%20agency%20has%20flagged%20%E2%80%9Cinscriptions,still%20at%20the%20analysis%20stage.

% \subsubsection{Non-Technical Challenges}
\smallskip
\noindent\textbf{User-friendliness and accessibility.} Although minting and transferring Bitcoin inscriptions can be done directly on Bitcoin, for many users, especially those not deeply familiar with blockchain technology, accessing and interacting with Bitcoin inscriptions may require specialized wallets and marketplaces. Most Ethereum users utilize wallets and tools optimized for interacting with ERC standards, which may not be compatible with Bitcoin inscriptions. This necessitates the use of Bitcoin-specialized wallets (e.g., Gamma.io \cite{gamma2024ordinals}, Ordinals Wallet \cite{ordinalswallet2024}) or third-party services (e.g., UniSat \cite{unisatwallet2024}, Magic Eden \cite{magiceden2024}), which is a significant educational gap for users transitioning from Ethereum to Bitcoin inscriptions. In addition, users accustomed to Ethereum's transaction speed and fee structure may find Bitcoin's inscription system less efficient and economical.

\smallskip
\noindent\textbf{Increased transaction costs.} 
Bitcoin transaction costs are usually influenced by several factors, such as the size of transactions, network congestion, and priority of transactions \cite{easley2019mining}. Inscriptions often involve embedding additional data into a Bitcoin transaction. The larger and more complex this data, the more space it occupies in a block, driving up transaction fees. As inscriptions become more popular, they contribute to network congestion, further elevating the costs. Therefore, users might become more selective in creating inscriptions due to higher fees. Cost barriers will lead users to perceive inscribing on Bitcoin as most appropriate for artworks that are either small in size or of high value.

\subsection{Opportunities}
% 各大铭文项目

\smallskip
\noindent\textbf{Inscription derivative protocols.} After the surge in popularity of Bitcoin inscriptions, the trend quickly spread to other public blockchains, leading to the emergence of numerous inscription derivative protocols and imitation projects. These innovative projects and protocols have further enriched the inscription ecosystem. We categorize mainstream inscription protocols based on their respective public blockchains, analyzing and comparing the inscription protocols on each chain (shown in Table \ref{tab:inscription_projects}).

As discussed before, the BRC-20 protocol faces limitations with only four-letter tokens and susceptibility to front-running attacks. To improve BRC-20, ARC-20 \cite{arc20tokens2024} removes the four-character limit, allowing for more diverse gameplay. A unique project within this framework is ``Realm" \cite{realmMetaverse2024}, where each registered entity is a prefix text, ultimately owning the pricing rights to all suffixes. Rune \cite{luminex2024rune}, proposed by Ordinals founder Casey, is designed to issue Fungible tokens by inserting token data directly into UTXO scripts. Rune's implementation is similar to ARC20, while it includes the token quantity in the script data, making it more legitimate. In addition, RGB \cite{rgbblueprint2024} represents an ultimate scaling solution, turning smart contract states into concise proofs inscribed in BTC UTXO output scripts. RGB offers low transaction costs and high scalability. It is considered a Layer2 for BTC, leveraging BTC's security for smart contracts.

For other blockchains, DRC-20 \cite{drc20standard2024} on Dogecoin (PoW chain) is similar to BRC-20 but popular due to low transaction costs and strong meme appeal. Ethscriptions protocol \cite{ethscriptions2024introduction} on the Ethereum chain introduced ``dumb contracts", bringing functionality and practicality. ASC-20 \cite{avaxmarket2024overview}, BSC-20 \cite{bnbchaininscriptions2024}, PRC-20 \cite{prc20contract2024} on other EVM-compatible chains also enable inscription and index building for projects like AVAL, BNBS, and POLS. SOLs \cite{magicedenunlocking2024} on Solana began in November 2023, with a total of 21,000,000 inscriptions. The main focus is on NFTs, with indexing based on image or file hashes. SFNs \cite{sfnsdefibox2024} is recognized as the first inscription on the EOS chain. It adopts the AtomicAsset standard \cite{atomicassetscontract2024}, which provides a powerful feature set while minimizing unnecessary complexity and focusing on RAM efficiency. In addition, MRC-20 \cite{coinlivemove2024} powered by the Move blockchain offers an efficient and user-friendly approach to token management. APT-20 \cite{apt20protocol2024} is an experimental token standard that facilitates the creation and transfer of fungible tokens through the Ordinals protocol on the Aptos blockchain.

\vspace{-0.1in}
\begin{longtable}{
 >{\centering\arraybackslash}m{1.25cm}
 >{\raggedright\arraybackslash}m{2.25cm}
 >{\raggedright\arraybackslash}m{1.5cm}
 >{\raggedright\arraybackslash}m{1.9cm}
 >{\raggedleft\arraybackslash}m{1.75cm}
 >{\raggedleft\arraybackslash}m{3.5cm}}
\caption{Comparative Analysis of Inscription Derivative Protocols} \label{tab:inscription_projects} \\
\renewcommand{\arraystretch}{1.5} % Increase cell height
\setlength{\tabcolsep}{10pt} % Increase cell padding
\endfirsthead

\multicolumn{6}{c}%
{{\bfseries \tablename\ \thetable\ -- continued from previous page}} \\
\endhead

\endfoot

\endlastfoot

\rowcolor[HTML]{C0C0C0}
\textbf{\color{white}Icon} & \textbf{\color{white}Protocol} & \textbf{\color{white}Chain} & \textbf{\color{white}Inscription} & \textbf{\color{white}Total Amount} & \textbf{\color{white}Feature Description} \\ 
\includegraphics[width=1cm, height=1cm]{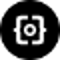} & BRC20  \cite{brc20whitepaper}               & BTC                 & Ordi             & 21,000,000            & First Bitcoin-based inscription protocol     \\ \hline
\includegraphics[width=1cm, height=1cm]{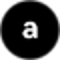} & ARC20   \cite{arc20tokens2024}              & BTC                 & Atom             & 21,000,000            & Removes the four-character limit        \\ \hline
\includegraphics[width=1cm, height=1cm]{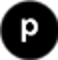} & Rune  \cite{luminex2024rune}               & BTC                 & Pipe              & 2,100,000             & Embeds token data in Bitcoin's UTXO          \\ \hline
\includegraphics[width=1cm, height=1cm]{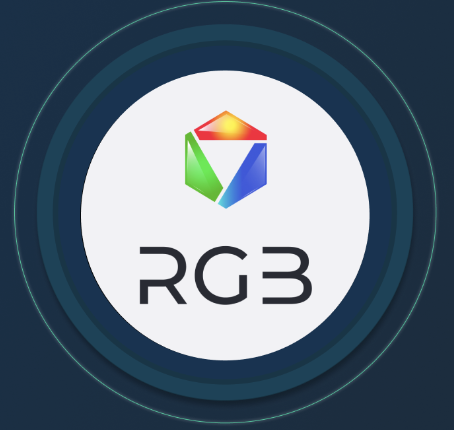} & RGB  \cite{rgbblueprint2024}               & BTC                 & RGB              & 2,100,000             & Converts smart contract states into concise proofs          \\ \hline
\includegraphics[width=1cm, height=1cm]{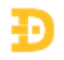} & DRC20    \cite{drc20standard2024}             & Doge                & Dogi             & 21,000,000            & Similar to BRC-20, popular for low costs          \\ \hline
\includegraphics[width=1cm, height=1cm]{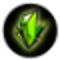} & Ethscription   \cite{ethscriptions2024introduction}              & ETH                 & Eths      & 21,000,000            & Introduces "dumb contracts"         \\ \hline
\includegraphics[width=1cm, height=1cm]{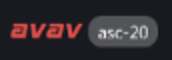} & ASC20   \cite{avaxmarket2024overview}              & Avax                & Avav             & 21,000,000            & EVM-compatible chain protocol         \\ \hline
\includegraphics[width=1cm, height=1cm]{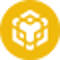} & BSC20   \cite{bnbchaininscriptions2024}              & BSC                 & Bnbs             & 21,000,000            & EVM-compatible chain protocol          \\ \hline
\includegraphics[width=1cm, height=1cm]{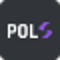} & PRC20  \cite{prc20contract2024}               & Polygon             & Pols             & 2.1e+15               & EVM-compatible chain protocol          \\ \hline
\includegraphics[width=1cm, height=1cm]{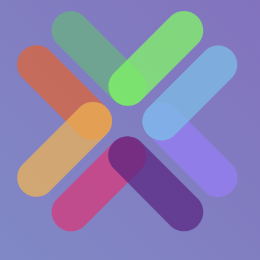} & AtomicAsset   \cite{atomicassetscontract2024}              & EOS                 & Snfs      & 21,000,000           & Efficient and powerful feature sets        \\ \hline
\includegraphics[width=1cm, height=1cm]{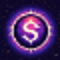} & SPL20  \cite{magicedenunlocking2024}               & Solana              & Sols             & 21,000,000            & NFTs with indexing based on file hashes         \\ \hline
\includegraphics[width=1cm, height=1cm]{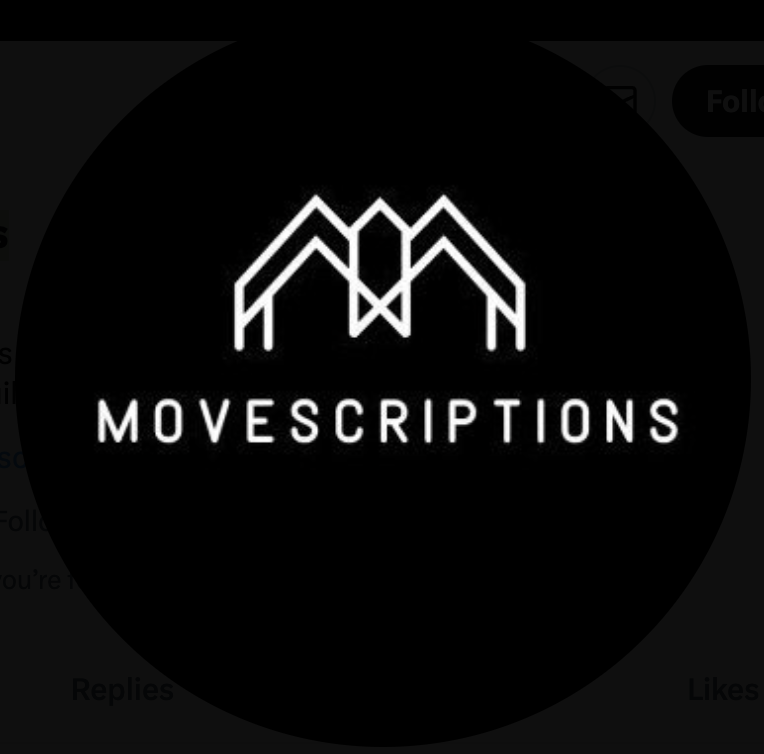} & MRC20   \cite{coinlivemove2024}              & Move              & Moves             & 21,000,000            & User-friendly approach to token management           \\ \hline
\includegraphics[width=1cm, height=1cm]{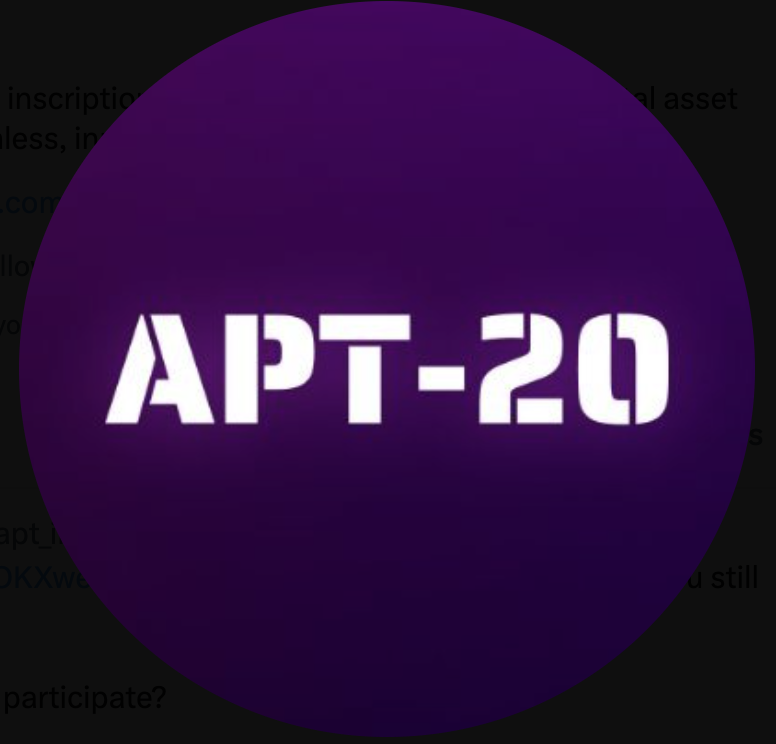} & APT20   \cite{apt20protocol2024}              & Aptos               & Apts             & 21,000,000            & Based on Ordinals protocol           \\ \hline 

\end{longtable}

\smallskip
\noindent\textbf{Bitcoin Layer2 solution.}
BTC Layer2 is a layer above the BTC network, primarily aimed at solving issues of insufficient transaction throughput, high transaction costs, and scalability challenges in the BTC network \cite{sguanci2021layer}. In simple terms, Layer1 refers to the Bitcoin public chain. To address the throughput issues of the BTC network and avoid high fees, transactions can be processed on Layer2 and then the results returned to Layer1, thereby reducing network pressure on the Bitcoin network. As the volume of inscription transactions on the Bitcoin network continues to grow, maintaining the stability of the settlement layer (Layer1) while encouraging innovation in the upper layer (Layer2) is the current and future main direction of development.

The core concept of ZK Rollups \cite{neiheiser2023practical} is to bundle multiple transactions into a single transaction on the Bitcoin blockchain. This process uses zero-knowledge proofs to verify bundled transactions without exposing details. The current versions of ZK rollups have raised concerns about centralization, as they primarily rely on centralized sequencers. In many existing implementations, a single entity is responsible for aggregating transactions, generating validity proofs, and submitting batch data to the Bitcoin network. This places considerable trust in the sequencers. In the future, a hybrid model combining multiple types of provers to accommodate different use cases may emerge. For example, a threshold scheme \cite{thibault2022blockchain} can distribute power to a dynamic group of sorting nodes based on equity or rotation; An opcode scheme \cite{mccorry2021sok} can enable bidirectional transfer of Sats and assets between the Bitcoin base layer and ZK rollups.

In addition to ZK rollups, other Layer2 technologies are also maturing. Two notable examples are Rootstock and Stacks. Rootstock (RSK) uses merged mining to ensure security comparable to Bitcoin and achieves expansion, efficiency, and advanced functionalities \cite{lerner2015rsk}. Merged mining allows Bitcoin miners to process and validate BTC and RSK transactions within the same block. In this mode, miners can mine simultaneously on the parent chain (larger blockchains like Bitcoin) and the child chain (smaller blockchains like RSK). By leveraging the computational power of the more powerful parent chain, the smaller chain gains additional security. Despite progress, RSK still faces challenges. RSK has difficulty attracting enough users, and the complexity and novelty of its merged mining mechanism also pose risks. Stacks is a Layer2 smart contract protocol designed specifically for Bitcoin, aiming to bring decentralized applications and smart contract functionalities to the Bitcoin ecosystem \cite{gudgeon2020sok}. Stacks introduces decentralized mining/bridging with Bitcoin. However, bottlenecks such as user experience, fees, and network effects may still be barriers to its broader adoption.

In 2023, the payment channels on the Lightning Network (LN) circulated over 5400 BTC, valued at over 230 million USD \cite{cointelegraphln2023}. LN micropayment channels establish a relationship between two parties, allowing them to continuously adjust balances without broadcasting each transaction to the blockchain \cite{poon2016bitcoin}. This method delays broadcasting the total balance between the two parties to a future point in time, effectively processing the total balance in one transaction. This approach allows financial relationships to exist without trust in the other party, free from the risk of default. The Taproot asset protocol \cite{LightningLabs2023}, announced in November 2023, further supports asset issuance through LN and provides customizable asset destruction features. Although high-fee events caused by network congestion have exposed LN's ongoing scalability limitations, LN solutions are expected to continue to be integrated into a wider range of applications as decentralized payment channels.

\smallskip
\noindent\textbf{Interoperability for Bitcoin inscriptions.}
Bitcoin inscriptions currently face limitations in interoperability and lack effective liquidity mechanisms. Establishing the infrastructure to provide liquidity for Bitcoin inscriptions, such as interoperability mechanisms~\cite{wang2023exploring}, is one of the key directions to propel the development of the Bitcoin inscription ecosystem. Interoperability techniques allow Bitcoin inscriptions to be traded or utilized in various blockchains. This opens up a larger market for these assets, as they can be accessed by users on different blockchain platforms, not just those on the Bitcoin network.

In 2023, Ordinals Market and Bitcoin Miladys jointly released the BRC721E standard \cite{maticz2024}. This standard enables the migration of verifiable ERC721 NFTs from Ethereum to Bitcoin Ordinals. During the bridging of NFTs, BRC721E encodes NFT data directly into a burn transaction. This burn transaction also acts as a request for a Bitcoin chain inscription, specifying a Bitcoin address to receive the inscription. It's important to note that the burn transaction is irreversible, which means that currently, it's not possible to convert an Ordinal back into an ERC-721 NFT. Therefore, we consider this reversibility a potential research direction for the future.

In addition, several cross-chain bridging protocols support the transfer of BRC-20 assets across chains, such as MultiBit \cite{multibitexchange2024}, TeleportDAO \cite{teleportdao2024}, and SoBit Bridge \cite{Sobit2023}. MultiBit is a cross-chain protocol between BRC20 and ERC20. It enables the transfer of tokens between BRC20 and ERC20. Similar to MultiBit, TeleportDAO has established a cross-chain Ordinals market, supporting the transfer and trading of assets between BTC and Polygon. SoBit Bridge allows users to transfer their BRC20 assets to Solana, enabling the creation of equivalent tokens on the Solana blockchain. Additionally, Allins \cite{allins2023} is a decentralized exchange (DEX) that employs an automated market maker (AMM) \cite{xu2023sok} method to establish asset liquidity for inscriptions. Allins encapsulates the inscription scripts of different chains in its unique virtual machine and continuously updates assets through an indexer. Leveraging smart contract-based liquidity pools, users can buy and trade inscriptions via AMM and earn income from inscription assets in the Farming market.

% \qw{At last I think it is better to add a table to summarise all mentioned Bitcoin-related standards, such as BIP141,341,341,342,..., BRC721E, ERC721, ERC20, ..., PRC20, (including all standard mentioned in this work, and all other platform's inscription standards), etc.} 

%-----------------------------------------------------
%Sec 6
\section*{Dessert: Very Short Summary}
In this report, we present an exploration of Bitcoin inscriptions, dissecting their technological roots. We compare Bitcoin inscriptions with traditional NFTs, revealing unique characteristics and broad use cases. Despite several challenges, we still highlight significant opportunities provided by Bitcoin inscriptions, especially for derivative protocols, Layer2 solutions, and interoperability techniques.

\section*{Sides: Bitcoin-related Standards}

We further present a table that includes a list of Bitcoin-related standards mentioned in this report, providing a quick reference for readers.

\begin{longtable}{|L{2.5cm}|L{6cm}|L{2.5cm}|L{1cm}|}
\hline
\rowcolor{gray!50}
\textbf{Standard and Proposal} & \textbf{Title} & \textbf{Blockchain Platform} & \textbf{Year} \\
\endfirsthead

\multicolumn{4}{c}%
{{\bfseries Table \thetable\ Continued from previous page}} \\
\hline
\rowcolor{gray!50}
\textbf{Standard and Proposal} & \textbf{Title} & \textbf{Blockchain Platform} & \textbf{Year} \\
\endhead

\hline \multicolumn{4}{|r|}{{Continued on next page}} \\ \hline
\endfoot

\hline \hline
\endlastfoot

\hline
\href{https://bips.xyz/114}{BIP 114} & Merkelized Abstract Syntax Tree & Bitcoin & 2016 \\
\hline
\href{https://github.com/bitcoin/bips/blob/master/bip-0141.mediawiki}{BIP 141} & Segregated Witness (Consensus layer) & Bitcoin & 2017 \\
\hline
\href{https://github.com/bitcoin/bips/blob/master/bip-0340.mediawiki}{BIP 340} & Schnorr Signatures for secp256k1 & Bitcoin & 2020 \\
\hline
\href{https://github.com/bitcoin/bips/blob/master/bip-0341.mediawiki}{BIP 341} & Taproot: SegWit Version 1 Spending Rules & Bitcoin & 2020 \\
\hline
\href{https://github.com/bitcoin/bips/blob/master/bip-0342.mediawiki}{BIP 342} & Tapscript & Bitcoin & 2021 \\
\hline
\href{https://ordinal-protocol.gitbook.io/ordinal-protocol/}{Ordinal} & Ordinal Protocol Documentation & Bitcoin & 2023 \\
\hline
\href{https://maticz.com/brc-721e-token-standard}{BRC-721E} & BRC-721E Token Standard & Bitcoin & 2023 \\
\hline
\href{https://www.blog.luminex.io/luminex-introduces-rune-issuance-tool}{Rune} & Rune Issuance Tool & Bitcoin & 2023 \\
\hline
\href{https://rgb-org.github.io/}{RGB} & RGB Blueprint: Scalable \& Confidential Smart Contracts & Bitcoin & 2023 \\
\hline
\href{https://docs.atomicals.xyz/arc20-tokens}{ARC-20} & ARC-20 Tokens - Atomicals Guidebook & Bitcoin & 2023 \\
\hline
\href{https://docs.drc-20.org/}{DRC-20} & DRC-20 Standard & Bitcoin & 2023 \\
\hline
\href{https://docs.ethscriptions.com/overview/introducing-ethscriptions}{Ethscriptions} & Introducing Ethscriptions & Ethereum & 2023 \\
\hline
\href{https://docs.avaxmarket.xyz/introduction/overview}{ASC-20} & Overview - ASC-20 AVAX Market & Avalanche & 2023 \\
\hline
\href{https://www.gate.io/learn/articles/understanding-bnb-chain-inscriptions/1473}{BSC-20} & Understanding BNB Chain Inscriptions & BSC & 2023 \\
\hline
\href{https://docs.platon.network/docs/en/PRC20_contract}{PRC-20} & PRC-20 Contract & Polygon & 2023 \\
\hline
\href{https://github.com/pinknetworkx/atomicassets-contract}{AtomicAsset} & AtomicAssets Smart Contract Repository & EOS & 2023 \\
\hline
\href{https://help.magiceden.io/en/articles/8615097-unlocking-the-future-inscriptions-on-solana}{SPL-20} & Unlocking the Future: Inscriptions on Solana & Solana & 2023 \\
\hline
\href{https://www.coinlive.com/news/what-s-going-on-with-move-smart-inscription-mrc20}{MRC-20} & What's Going On With MOVE: Smart Inscription (MRC20) & Move & 2023 \\
\hline
\href{https://apt-20.com/}{APT-20} & APT-20 Protocol & Aptos & 2023 \\
\hline
% ... Add all other rows in a similar fashion
\end{longtable}

%================================================
\bibliographystyle{splncs04}
\normalem
\bibliography{bib}
%================================================

\appendix

\end{document}